\documentclass[aps,prl,showpacs,twocolumn,groupedaddress,amsmath]{revtex4}

\usepackage{graphicx}

\begin{document}
\title{Competition of Percolation and Phase Separation in a Fluid of Adhesive Hard Spheres}
\author{Mark A.~Miller and Daan Frenkel}
\affiliation{FOM Institute for Atomic and Molecular Physics, Kruislaan 407,
1098 SJ Amsterdam, The Netherlands}
\date{\today}

\begin{abstract}
Using a combination of Monte Carlo techniques, we locate the
liquid--vapor critical point of adhesive hard spheres.  We
find that the
critical point lies deep inside the gel region of the phase
diagram. The (reduced) critical temperature and density are
$\tau_c=0.1133\pm0.0005$ and $\rho_c=0.508\pm0.01$. We
compare these results with the available theoretical predictions.
Using a finite-size scaling analysis, we verify that the critical
behavior of the adhesive hard sphere model is consistent with that
of the 3D Ising universality class.
\end{abstract}


\maketitle

The structure of a simple liquid is well described by
that of a system of hard spheres at the same effective density. To
a good approximation, the effect of attractive interactions on the
liquid structure  can be ignored. This feature of simple liquids
is implicit in the Van der Waals theory for the liquid--vapor
transition, and has been made explicit in the highly successful
thermodynamic perturbation theories for simple
liquids~\cite{perturbation}. The
perturbation approach becomes exact as the range of the attractive
interaction tends to infinity while its integrated strength
remains constant~\cite{longrange}. We refer to this limit
as the `Van der Waals' (VDW)
limit \cite{vdwnote}.  Conversely, as
the attractive forces become shorter-ranged and stronger, the
perturbation approach is likely to break down. Fluids with strong,
short-ranged attraction (so-called `energetic'
fluids~\cite{energetic}) are of growing importance in the area of
complex liquids. For example, short-range attractions are thought
to be responsible for the transition from a `repulsive' to an
`attractive' glass~\cite{energetic}, which has recently been observed
experimentally in PMMA (polymethylmethacrylate) dispersions \cite{pmma}.

In this Letter, we consider a model system that can be considered
as the prototypical energetic fluid: a fluid of adhesive hard
spheres (AHS).  Introduced in 1968 \cite{model}, the AHS model
is a reference system
for particles with short range attractions.  The pair potential
consists of an impenetrable core plus a surface adhesion term that
favors configurations where spheres are in contact.  At larger
separations, there is no interaction. The AHS model can be
considered as the `anti Van der Waals' limit.

Baxter showed~\cite{model} that the Percus--Yevick (PY) equation
can be solved analytically for adhesive hard spheres.  In fact, Baxter's
solution is often used to to analyze experimental results for systems as
diverse as silica suspensions \cite{silica}, copolymer
micelles \cite{micelles}, and the fluid phase of lysozyme
\cite{lysozyme}.

One important feature of the AHS model is that its phase diagram
contains a liquid--vapor coexistence
region~\cite{pathological}.
The PY equation offers different routes to estimate the location of
the liquid--vapor critical point. However, the `compressibility
route'~\cite{model} and the `energy route'~\cite{energyroute}
lead to estimates for the critical temperature that differ by some
20\%, while the estimates for the critical density differ by
almost a factor of three. For the analysis of experimental data, it
is important to know the location of the critical point more
accurately. The reason is that, upon cooling, adhesive hard
spheres can percolate to form a gel. Information about the
location of this percolation curve is available from
simulation~\cite{method,lee} and from analytical
estimates~\cite{chiew}. It is not clear, though, how
percolation interferes with the fluid--fluid phase separation. If
Baxter's estimate of the critical point were right, the critical
point would be on, or near, the percolation curve.  However,
if the estimate of Ref.~\cite{energyroute} were closer to
reality, then the fluid--fluid critical point would be deep
inside the gel phase. Clearly, this difference has implications
for the possibility to observe the fluid--fluid critical point in
systems with short-ranged attraction. Also for the (attractive)
glass transition, it makes a considerable difference whether or
not the transition line runs close to a liquid--vapor critical
point.

In spite of its obvious importance as a reference system for
`energetic' (as contrasted to `entropic') complex
liquids~\cite{energetic}, there exist, to our knowledge, no accurate
numerical estimates of the critical point of the AHS model. This
is not surprising, as computer simulations of the low-temperature
AHS model are notoriously difficult, due its propensity to form
large, even percolating, clusters. In this Letter, we report a
Monte Carlo (MC) simulation study that allows us to locate the AHS
critical point.

The AHS interaction is derived from a square well potential by
taking a limit in which the well becomes infinitesimally narrow at
the same time as becoming infinitely deep in such a way that the
integrated Boltzmann weight of bound configurations remains finite
\cite{model}.  The interaction is most
easily defined by the expression for the Boltzmann factor as a
function of pair separation, $r$:
\begin{equation}
\exp[-U(r)/kT] = \Theta(r-\sigma) +
\frac{\sigma}{12\tau}\delta(r-\sigma). \label{boltzmann}
\end{equation}
In Eq.~(\ref{boltzmann}), $U(r)$ and $kT$ are the formal pair
potential and thermal energy, respectively, while $\sigma$ is the
hard core diameter, which we henceforth use as the unit of length.
The step function $\Theta$ accounts for the hard sphere repulsion,
and the Dirac delta function introduces the surface adhesion.  The
parameter $\tau$ determines the strength of the adhesion and can
be interpreted as an effective temperature, or an `inverse
stickiness parameter.'

To compare the properties of the
AHS and VDW  models, we need a measure for the strength of the
attractive interactions that is meaningful in both limits. This is
best achieved by comparing the reduced second  virial
coefficients: $B_2^*\equiv B_2/B_2^{\rm HS}$,
where $B_2^{\rm HS}$ is the second virial coefficients of hard
spheres. For the AHS model, $B_2^*=1-1/(4\tau)$, and we use this
expression to define the $\tau$-parameter for the VDW model. In
the VDW limit, the free-energy density is given by
$f_{\rm VDW}(\rho)=f_{\rm HS}(\rho)-a\rho^2$ where $f_{\rm HS}$ is the
free energy density of a system of hard spheres at density $\rho$,
while the constant $a$ measures the strength of the attractive
forces. Hence, in the VDW limit, $\tau=kTB_2^{\rm HS}/(4a)$ is
proportional to the temperature.  We henceforth refer to $\tau$ as
the temperature parameter for both the VDW and AHS systems.

In the AHS model, particles may stick and form clusters.
At sufficiently low $\tau$, clusters may percolate
(span the system), mimicking the infinite clusters that form during
gelation in the real system.  Percolating clusters
pose serious problems for any simulation scheme that employs
volume changing moves (e.g.~constant pressure MC), since all
such moves are rejected as soon as percolating clusters appear.

To study the phase behavior of adhesive hard spheres, we therefore
used Grand Canonical MC (GCMC) simulation.  This technique
can be used beyond the percolation threshold.  Yet, even with
GCMC, equilibration of the system at low
temperatures is prohibitively slow. We therefore combine GCMC
with parallel tempering to speed up equilibration, and with
multiple histogram reweighting, to make optimal use of the
available simulation data.  All simulation techniques had to be
specifically adapted for the AHS model.

The fact that the attractive term in Eq.~(\ref{boltzmann}) is
infinitesimally narrow means that, in a simulation, the chance of
generating a bound configuration by random displacement of a
particle is vanishingly small.  Conversely, the probability of
breaking such a bond, once formed, is also negligible.
Consequently, conventional Metropolis sampling cannot be applied to
the AHS model.

We therefore employ a modification of the AHS-MC schemes described
in Ref.~\cite{method}, where single particle displacements
explicitly make and break up to three contacts with the test
particle simultaneously.  States with higher coordination numbers
can be reached indirectly through suitable combinations of moves.

In our GCMC simulations, we employ particle insertion
and removal steps with equal probability that together constitute
45\% of MC steps. We only insert and remove particles that are not
bound to others.

To speed up equilibration, we perform cluster translation moves
with probability 5\%.  A particle is chosen at random, and the
cluster to which it belongs is translated by a random amount in
each Cartesian direction up to a maximum that is inversely
proportional to the number of particles in the cluster.

To overcome the slow equilibration of large clusters at low $\tau$
or high density, we have used the parallel
tempering scheme of Geyer~\cite{tempering}.
Parallel tempering can be performed with replicas at different
temperatures, different chemical potentials, or combinations of
both.  We have found it most convenient to gather statistics
across the full density range for one temperature at a time, and
therefore chose to run a hierarchy of chemical potentials, $\mu$,
with the same value of $\tau$. Each replica attempts a
configuration exchange with one of its neighbors in $\mu$
(alternating between the higher and lower neighbor) every 200 MC
steps.  The acceptance probability for an exchange between
replicas $i$ and $j$ is ${\rm min}[1,(z_i/z_j)^{N_j-N_i}]$, where
$N_i$ is the number of particles in the current configuration of
replica $i$, and the activity $z$ is related to the chemical
potential by $z=\Lambda^{-3}\exp(\mu/kT)$, with $\Lambda$ the
thermal de Broglie wavelength.

The final computational tool needed to analyze the data is multiple
histogram reweighting \cite{multihist}.
The probability of observing the system in a state with $N$ particles and $B$
binary contacts between the particles at temperature $\tau$ and activity $z$
can be exactly decomposed into the form
\begin{equation}
p_{NB}(\tau,z) = \Omega_{NB}\tau^{-B}z^N/\Xi(\tau,z),
\label{multhist}
\end{equation}
where $\Omega_{NB}$ is an effective density of states, and $\Xi(\tau,z)$ is
the grand canonical partition function.  Knowledge of $\Omega_{NB}$ up to a
multiplicative constant is sufficient to calculate the distribution of
$N$ or $B$ at any set of conditions ($\tau,z$).  $\Omega_{NB}$ can be obtained
over a wide range of ($N,B$) by combining overlapping two-dimensional histograms
of $N$ and $B$ collected from simulations at different $\tau$ and $z$.

Simulations at sufficiently low $\tau$ show a range of activities
in which the density distribution is bimodal, indicating the
coexistence of a high- and a low-density fluid phase.  We studied
this coexistence at four different system sizes, labeled by the
length $L$ of the cubic simulation box.  For $L=5,6$ and $8$, the
simulations consisted of $10^6$ equilibration blocks and $10^7$
acquisition blocks, where a block is $L^3/\sigma^3$ MC trial moves,
each of which may be a particle displacement,
particle insertion/removal, cluster translation, or replica
exchange.  At each temperature, seven or eight parallel replicas
were used, with the range of activities chosen to span the
coexisting densities.  At $L=10$, the numbers of equilibration and
acquisition blocks had to be lowered to $2.5\times10^5$ and
$2.5\times10^6$, respectively, to obtain results in a reasonable
computational time.

\begin{figure}
\includegraphics[width=80mm]{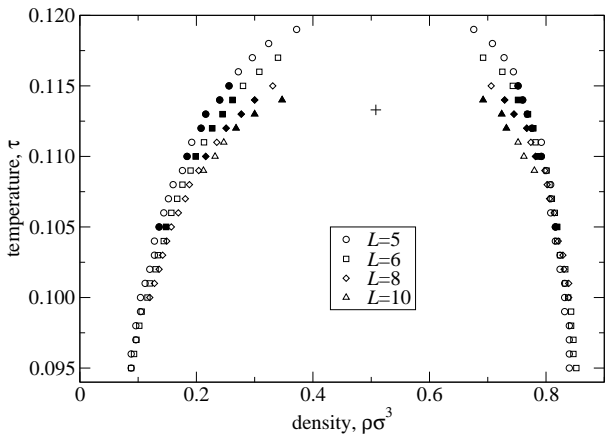}
\caption{Coexistence curve of the AHS model from GCMC
using four box lengths, $L$.  Solid
symbols denote points actually simulated; open symbols were
obtained by histogram reweighting. Error bars
are omitted for clarity. The cross denotes the estimated critical
point at $\tau=0.1133$, $\rho=0.508$. \label{coexistence} }
\end{figure}

The $(N,B)$ histograms for different $\tau$ and $z$ were combined
for each system size separately.  The coexisting densities were
then obtained as a function of temperature by reweighting the
histograms to find the activity at which the two density peaks
have equal height. Figure \ref{coexistence} shows the resulting
coexistence curves. Strong finite size effects are immediately
visible.  The curves obtained from different system sizes coincide
at sufficiently low temperature, but deviate significantly from
each other as the critical region is approached.

The size dependence of the coexistence distributions can be used
to locate the critical point more accurately using the approach
developed by Bruce and Wilding
\cite{universality,ljfluid}.  This method uses the fact
that, within a universality class, the critical distribution of
the order parameter is invariant up to a rescaling of the
order parameter.  As the AHS interactions are short-ranged,
the fluid--fluid critical point of this model is expected
to belong to the 3D Ising universality class.  The critical
distribution is known to high precision from studies of lattice
systems, and accurate analytic fits are available
\cite{ising}.

Due to the absence of particle--hole symmetry
in off-lattice models, the distribution of the density in the AHS
model is not quite symmetric about its mean.  Symmetry can be
restored by accounting for the mixed character of the scaling
fields.  The appropriate order parameter is then no longer the
pure particle density $\rho=N/L^3$, but includes a contribution
from the energy density, which in the AHS model can be taken as
$u=-B/L^3$.  The particle and energy density operators are
replaced by the linear combinations \cite{universality}
\begin{equation}
{\cal E}=\frac{u-r\rho}{1-sr} \qquad {\cal M}=\frac{\rho-su}{1-sr},
\end{equation}
where $s$ and $r$ are system-dependent field mixing parameters that are
identically zero for models with Ising symmetry.  Precisely at criticality,
the distribution of $\cal M$ in a system of sufficiently large linear length
$L$ takes on the universal form
$p_L({\cal M}) = p^*(a_L[{\cal M}-{\cal M}_c])$,
where ${\cal M}_c=\langle{\cal M}\rangle$ evaluated at the critical temperature,
and $a_L\propto L^{\beta/\nu}$ with $\beta$ and $\nu$ critical exponents.

\begin{table}
\caption{\label{critical} Size-dependent properties for critical
point determination. $\tau_a$ is the apparent critical
temperature, at which the distribution of $\cal M$ (with
field-mixing parameter $s$) collapses onto the universal form
$p^*(x)$. The remaining three columns refer to properties at the
proposed critical temperature of $\tau^{\rm sim}_c=0.1133$:
$\rho_c$ is the mean density, $z_c$ is the activity, and $a_L$
scales $p(a_L{\cal M})$ to have unit variance. }
\begin{ruledtabular}
\begin{tabular}{llllll}
$L$ & $s$ & $\tau_a$ & $\rho_c$ & $z_c$ & $a_L$ \\
\hline
5  & 0.04 & 0.1130 & 0.499 & 0.08809 & 3.830 \\
6  & 0.04 & 0.1134 & 0.506 & 0.08762 & 4.036 \\
8  & 0.02 & 0.1135 & 0.513 & 0.08727 & 4.727 \\
10 & 0.02 & 0.1132 & 0.512 & 0.08723 & 5.362 \\
\end{tabular}
\end{ruledtabular}
\end{table}

\begin{figure}
\includegraphics[width=80mm]{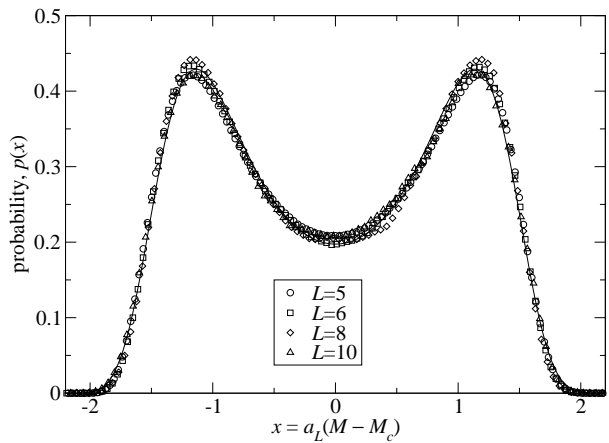}
\caption{Distribution of the order parameter at the proposed
critical temperature from GCMC at four
different box lengths, $L$.  Each curve has been scaled to
have unit variance.  The solid curve is the 3D Ising
critical distribution \cite{ising}. \label{distribution} }
\end{figure}

Table \ref{critical} lists the apparent critical temperature
$\tau_a$ and mixing parameter $s$ at which the distribution of
${\cal M}$ most closely matches $p^*(x)$ for the four system sizes
studied.  The fact that $s$ is small indicates that the
distribution of $\rho$ itself is nearly symmetric. The apparent
size-dependence of $s$ is probably---at least partly---due to the fact
that, despite the long simulations employed, statistical
fluctuations in the histograms are still significant. For small
$L$, $\tau_a$ is expected to show some size dependence due to
finite-size corrections to scaling \cite{ljfluid}. However, the
differences in $\tau_a$ shown in Table \ref{critical} are not
monotonic, and are again likely to be due to limited statistics.
Rather than attempting to extrapolate to the
infinite system limit, we therefore quote the critical parameters
that most closely reproduce the universal critical form, with
conservative error bars:
\begin{equation*}
\tau^{\rm sim}_c=0.1133\pm0.0005 \qquad \rho^{\rm sim}_c=0.508\pm0.01.
\end{equation*}
Figure \ref{distribution} shows the distribution of the order parameter at
the proposed $\tau^{\rm sim}_c$ (not the individual apparent critical
temperatures $\tau_a$) for all four system sizes.  The collapse of the data
onto the Ising distribution shows that the transition is
consistent with this universality class.

The size dependence of the scaling factor $a_L$ in Table~\ref{critical}
permits an
estimate of the ratio of the critical exponents $\beta$ and $\nu$.
A straightforward fit to the asymptotic form
$a_L\propto L^{\beta/\nu}$ yields
$\beta/\nu=0.50\pm0.04$, which is clearly compatible with the
Ising value $\beta/\nu=0.52$.

\begin{figure}
\includegraphics[width=80mm]{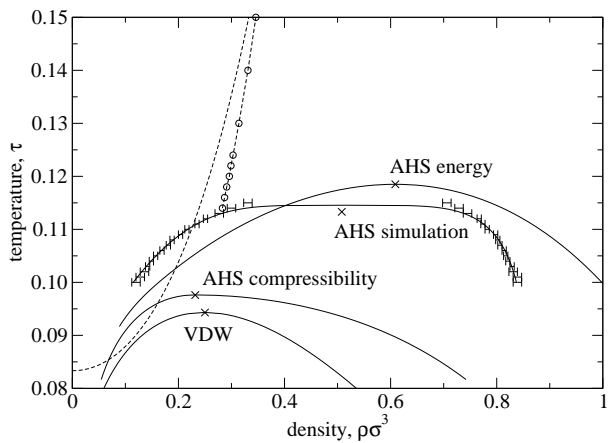}
\caption{Solid lines: coexistence curves for the VDW limit,
the AHS PY compressibility and energy routes, and AHS simulation
with $L=8$.
For the latter, the line merely guides the eye between the data points.
Dashed lines: percolation threshold from PY theory
and simulation (line with circles).
\label{py}
}
\end{figure}

We can now compare our numerical results with various theoretical
predictions. Below, we list the estimates based on the PY
compressibility \cite{model} and energy \cite{energyroute}
routes, and the estimate based on the VDW (mean-field)
expression \cite{meanfield}:
\begin{align*}
\tau^{\rm PYc}_c&=0.0976 &\quad \rho^{\rm PYc}_c&=0.232\\
\tau^{\rm PYe}_c&=0.1185 &\quad \rho^{\rm PYe}_c&=0.609\\
\tau^{\rm VDW}  &=0.0943 &\quad \rho^{\rm VDW}  &=0.250.
\end{align*}
The coexistence curves for the two PY routes can be obtained
numerically using the analytical expressions \cite{analytic}
for the chemical potential and pressure.
These theoretical curves, together with the
mean-field result and the  simulation data at $L=8$
are shown in Fig.~\ref{py}. The simulation results lie between the
two PY predictions, but are clearly closer to those of the energy
equation.

Also plotted in Fig.~\ref{py} is the percolation threshold.  The
theoretical result \cite{chiew} corresponds to the PY estimate
of the points where the cluster size diverges.  In simulations, we
define the percolation threshold as the locus of points were the
probability of observing a percolating cluster in a canonical
simulation is 50\%.  This definition is relatively size
independent, and our results with $N=500$ particles are consistent
with the more elaborate analysis of Lee \cite{lee}.
Both the theoretical and simulated percolation lines
clearly show that the critical point of the
fluid--fluid transition lies well within the percolated part of
the phase diagram.

The phase diagram reported in Fig.~\ref{py} is
likely to be representative of that of real colloidal systems with
short-ranged attraction at the same value of $\tau$. If so, our
findings suggest that the attractive glass
transition~\cite{pmma,attractive} is relatively far removed from the
fluid--fluid critical point.

The work of the FOM Institute is part of the research program of
FOM and is made possible by financial support from the Netherlands
organization for Scientific Research (NWO). We thank
Dr.~R.~P.~Sear for the mean-field data in Fig.~\ref{py}, and gratefully
acknowledge insightful discussions with
Prof.~W.~C.~K.~Poon and Dr.~N.~Kern.


\begin{thebibliography}{22}
\bibitem{vdwnote}
In our definition of the VDW limit
we assume that the free energy density of hard spheres is known
exactly.  Van der Waals used an approximate expression for this term.
\end{thebibliography}
\end{document}